\documentclass[%
twocolumn,
superscriptaddress,
amsmath,amssymb,
prl,
floatfix,
]{revtex4-2}

\usepackage{xcolor}
\usepackage{graphicx}
\usepackage{bm}
\usepackage{hyperref}
\usepackage{lineno}

\usepackage{amsfonts}

\usepackage[english]{babel}
\usepackage{amsthm}

\newtheorem{theorem}{Theorem}

\theoremstyle{definition}
\newtheorem{definition}{Definition}

\usepackage{bbold}
\usepackage{mathtools}

\begin{document}


\title{Turing-Completeness and Undecidability in Coupled Nonlinear Optical Resonators}

\author{Gordon H.Y. Li}
\affiliation{Department of Applied Physics, California Institute of Technology, Pasadena, CA 91125, USA}
\author{Alireza Marandi}
\email{marandi@caltech.edu}
\affiliation{Department of Applied Physics, California Institute of Technology, Pasadena, CA 91125, USA}
\affiliation{Department of Electrical Engineering, California Institute of Technology, Pasadena, CA 91125, USA}


\begin{abstract}
\vspace{3mm}
\noindent \textbf{Abstract:} Networks of coupled nonlinear optical resonators have emerged as an important class of systems in ultrafast optical science, enabling richer and more complex nonlinear dynamics compared to their single-resonator or travelling-wave counterparts. In recent years, these coupled nonlinear optical resonators have been applied as application-specific hardware accelerators for computing applications including combinatorial optimization and artificial intelligence. In this work, we rigorously prove a fundamental result showing that coupled nonlinear optical resonators are Turing-complete computers, which endows them with much greater computational power than previously thought. Furthermore, we show that the minimum threshold of hardware complexity needed for Turing-completeness is surprisingly low, which has profound physical consequences. In particular, we show that several problems of interest in the study of coupled nonlinear optical resonators are formally undecidable. These theoretical findings can serve as the foundation for better understanding the promise of next-generation, ultrafast all-optical computers. 
\end{abstract}

\maketitle
\section{Introduction}
Networks of coupled nonlinear optical resonators (CNORs) are a general class of optical systems combining nonlinear optical interactions, resonator or cavity feedback mechanisms, and linear couplings between resonators. CNORs have been well-studied in a variety of experimental platforms including pairs or dimers of coupled cavities~\cite{zhang2019electronically,tikan2021emergent,yelo2021self}, time-multiplexed cavities with time-delayed couplings~\cite{marandi2014network}, spatially-multiplexed arrays of cavities~\cite{parto2020realizing,tusnin2023nonlinear,okawachi2020demonstration}, and wavelength-multiplexed cavities with a synthetic frequency dimension~\cite{yuan2018synthetic}. They offer richer dynamics compared to their linear, traveling-wave, or single-cavity counterparts and have served as a fertile ground for exploring complex nonlinear dynamics such as solitons~\cite{yuan2023soliton,ji2024multimodality,gao2024observation}, phase transitions~\cite{roy2023non}, topological phenomena~\cite{roy2022topological,leefmans2024topological,leefmans2023cavity}, and non-Hermitian physics~\cite{roy2021nondissipative}. Furthermore, CNORs have technologically-important applications such as low-noise microwave generation~\cite{triscari2023quiet,kudelin2024photonic}, optical switching~\cite{scheuer2010all,chen2014tunable}, optical isolators~\cite{zhou2016pt,zhou2017optical}, and quantum state generation~\cite{yokoyama2013ultra,asavanant2019generation,larsen2019deterministic}. 

Here we are interested in better understanding the computational properties of CNORs. At an abstract level, CNORs can possess both strong nonlinearity and memory, which are the essential ingredients for computation. Indeed, CNORs have been demonstrated as physical Ising solvers for combinatorial optimization~\cite{wang2013coherent,marandi2014network,inagaki2016large}, as well as efficient hardware accelerators for deep learning~\cite{basani2024all} and neuromorphic computing~\cite{feldmann2019all,makinwa2023experimental}. However, previous experimental works treated CNORs as special-purpose or application-specific optical processors performing a single kind of computational task. In this work, we rigorously prove that CNORs are Turing-complete (see Theorem 1), which means that they can be used to compute any computable function, and thus possess much greater computational power than previously considered. This result is not obvious since CNORs are described by continuous-valued or analog field amplitudes instead of discrete-valued states, and have a completely dynamical memory instead of the static memory elements typically associated with Turing Machines.  

Turing-completeness also immediately implies that several properties of CNORs such as the existence of a steady-state or periodic oscillation are formally undecidable, i.e. there is no algorithm that can always correctly answer the associated decision problem (either in the affirmative or negative). This stems directly from the undecidability of the \textit{Entscheidungsproblem} (or Halting Problem)~\cite{turing1937on}, which we now interpret in the context of CNORs. Therefore, we may add CNORs to the rapidly growing list of physical theories that exhibit undecidable properties, e.g. classical mechanical systems~\cite{fredkin1982conservative,moore1990unpredictability,da1991undecidability}, fluid dynamics~\cite{tao2016finite,cardona2021constructing}, quantum many-body systems~\cite{cubitt2015undecidability,bausch2021uncomputability,shiraishi2021undecidability}, quantum field theory~\cite{komar1964undecidability,tachikawa2023undecidable}, and quantum gravity~\cite{geroch1986computability}.  

The general strategy for our mathematical proof is to show by explicit construction how any Turing Machine can be exactly simulated by an associated CNOR. This explicit construction also allows us to conveniently place an upper bound on the minimum hardware complexity needed for Turing-completeness in CNORs. If the minimum complexity needed for Turing-complete CNORs is high, then we could just simply agree to avoid such niche situations and hence also avoid vexing issues of undecidability. On the other hand, if the minimum complexity is low, then we should expect the study of CNORs (and ultrafast optical science more broadly) to be rife with examples of undecidable problems. Interestingly, we show that the latter is true and that the minimum complexity needed for Turing-complete CNORs is well-within current experimental capabilities. Therefore, we cannot easily dismiss issues of undecidability and must seriously confront what can and cannot be logically reasoned regarding CNORs. 

\section{Results}
\subsection{Computational models}
We begin by defining our computational models, then give a detailed proof of the main Turing-completeness result, and point out some interesting physical consequences for the study of CNORs. Finally, we discuss the limitations and practicalities associated with our results. 

\begin{figure}[b]
\includegraphics[width=\linewidth]{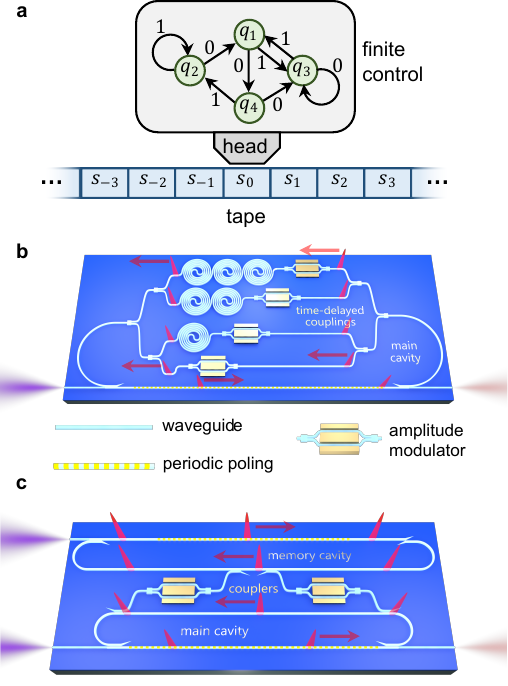}
\caption{\textbf{Computational models.} (a) A Turing Machine (TM) consists of a bi-infinite tape of symbols, a head that can read/write symbols and move left/right, and a finite state control. A degenerate optical parametric oscillator network (DOPON) can be implemented using time-multiplexing in which short laser pulses act as independent nonlinear resonators that interact via either (b) intra-cavity time-delayed couplings or (c) coupled cavities.}
\label{fig:1}
\end{figure}

We follow the definition of a (one-tape) Turing machine given by Minsky~\cite{minsky1967computation} as shown conceptually in Fig.~\ref{fig:1}a. Other equivalent definitions of Turing machines and their generalizations are also possible. 
\begin{definition}[Turing Machine]
\label{definition1}
A \textit{Turing Machine} (TM) is a \textit{finite-state machine} with access to a linear \textit{tape} that extends infinitely in both left and right directions. The tape consists of a sequence of cells each printed with a single symbol from a finite tape alphabet $S=\{s_1,\ldots,s_{n}\}$. A movable \textit{head} is situated on some cell and can read/write symbols on that cell. The current symbol read by the head is the input symbol to the finite-state machine with internal states $Q=\{q_{1},\ldots,q_{m}\}$. The TM is described by three (partial) functions $G:Q\times S\nrightarrow Q$, $F:Q\times S\nrightarrow S$, and $D:Q\times S\nrightarrow\{0,1\}$, denoting the updated state, new symbol to write, and direction to move the head, respectively. We take `0' to mean `move left' and `1' to mean `move right'. At each time step, the machine starts in some state $q_{i}$, reads the symbol $s_{j}$ written on the cell under the head, prints there the new symbol $F(q_{i},s_{j})$, moves one cell to the left or right according to $D(q_{i},s_{j})$, and then enters the new state $G(q_{i},s_{j})$.$\quad\square$
\end{definition}

Without loss of generality, henceforth we consider just a binary tape alphabet $S=\{0,1\}$ where `0' is the special \textit{blank} tape symbol. Importantly, the starting configuration of the tape can only contain a finite number of non-blank symbols. In the following, we consider a type of CNOR based on networks of time-multiplexed degenerate optical parametric oscillators with dissipitive couplings. These are the most well-studied and relevant class of CNORs for computing applications and have a wide range of possible experimental platforms including free-space optical systems~\cite{marandi2014network}, optical fiber networks~\cite{inagaki2016large}, and photonic integrated circuits~\cite{gray2024large}.

\begin{definition}[Degenerate Optical Parametric Oscillator Network] A \textit{Degenerate Optical Parametric Oscillator Network} (DOPON) is described by $N$ optical pulses with amplitudes $x_{i}\in\mathbb{R}$ for $i=1,2,\ldots,N$ that evolve in time according to the dynamical system in Eq.~\ref{eq:1}
\begin{equation}
x_{i}(t+1)=\rho\left[x_{i}(t)\right]+\sum_{j=1}^{N}J_{ij}(t)\cdot x_{j}(t)
\label{eq:1}
\end{equation}
where $t$ is the discrete time step, $J_{ij}(t)$ is the coupling weight between the $i^{\mathrm{th}}$ and $j^{\mathrm{th}}$ pulses at time step $t$, and $\rho:\mathbb{R}\rightarrow\mathbb{R}$ is an anti-symmetric (odd) function $\rho(-x)=-\rho(x)$ representing saturable parametric gain (i.e. linear for small signal values, and saturated for large signal values) in each optical resonator as shown in Eq.~\ref{eq:2}
 \begin{equation}
     \rho(x)=\begin{cases}
        1\ ,&\ \mathrm{if}\ x\geq1\\
        g(x)\ ,&\ \mathrm{if}\ 3/4< x< 1\\
        x\ ,&\ \mathrm{if}\ 0\leq x\leq 3/4\
    \end{cases}\\  
    \label{eq:2}
 \end{equation}
 where $g(x)$ is any continuous monotonically-increasing function defined over the domain $x\in[3/4,1]$.$\quad\quad\quad\quad\square$
\label{definition2}
\end{definition}

We consider two possible designs for DOPONs, which can both be operated in a way that obeys Eq.~\ref{eq:1}. The first implementation, shown schematically in Fig.~\ref{fig:1}b is based on a single main cavity containing multiple short laser pulses. Each pulse occupies one of equally-spaced time bins that act as independent optical resonators experiencing the nonlinear optical parametric gain mechanism (e.g., provided by periodically-poled lithium niobate waveguides~\cite{ledezma2022intense}). The pulses can be coupled using intra-cavity optical delay lines with delays matching the temporal separation of pulses, shown as a multi-arm Mach-Zehnder interferometer in which coupling weights are set using amplitude modulators (e.g. electro-optic modulators~\cite{wang2018integrated}). In this configuration, all-to-all coupling can be achieved for a main cavity containing $N$ pulses if there are $N$ optical delay lines in the intra-cavity interferometer. Therefore, the discrete time step $t$ represents the main cavity roundtrip number, as all pulses can be coupled after a single roundtrip. 

The second implementation, shown schematically in Fig.~\ref{fig:1}c is based on a main cavity containing $N$ equally-spaced pulses that is coupled to a second cavity containing $N+1$ pulses (with the same pulse spacing as the main cavity), which we call the memory cavity. The programmable couplers between the main and memory cavities allow the coupling between pulses to occur according to Eq.~\ref{eq:1}. To do this, the memory cavity is operated with a purely linear gain that compensates the roundtrip loss and the main cavity has a saturable gain as before. Coupling terms are accumulated one at a time with intermediate values stored in the memory cavity. Memory cavity pulses are only allowed to interfere with main cavity pulses after all coupling terms are accumulated. Therefore, in this configuration with all-to-all coupling, the discrete-time step $t$ represents $N+1$ roundtrips of the main cavity. Compared to the first implementation, the second implementation achieves a constant $\mathcal{O}(1)$ number of modulators and delay lines at the expense of more cavity roundtrips $\mathcal{O}(N)$ to achieve all-to-all coupling.

Some remarks about Definition~\ref{definition2}: (1) Previous studies~\cite{wang2013coherent} of DOPONs considered continuous-time dynamical systems. However, our use of discrete time in terms of the resonator roundtrip number is more realistic since the gain and couplings are experimentally implemented as lumped elements at fixed locations, and are not continuously distributed throughout the resonator. (2) A single degenerate optical parametric oscillator without couplings is described by $x(t+1)=\rho[x(t)]$ and has the normal form of a supercritical pitchfork bifurcation $x(t+1)=\alpha\cdot x(t)-x^{3}$. For $\alpha<1$, we say that the oscillator is \textit{below threshold} and the only stable fixed-point is the trivial solution $x=0$. For $\alpha>1$, the oscillator is \textit{above threshold} and has two stable fixed-point solutions $x=\pm\sqrt{\alpha-1}$. Previous studies of DOPONs considered only the normal form of the nonlinearity, however the inclusion of a saturable gain is more physical and ensures that the pulse amplitudes remain bounded. (3) Strictly speaking, there should be additional Langevin noise operators in Eq.~\ref{eq:1} when operating DOPONs below threshold, however for simplicity, we ignore small noise perturbations since we consider only above threshold DOPONs. Therefore, the trajectory of the DOPON is deterministic and completely specified by the coupling weights $\{J_{ij}(t)\}$ and initial conditions $x_{i}(0)$ for $i=1,2,\ldots,N$. We require that the couplings $\{J_{ij}(t)\}$ be known \textit{a priori}, i.e. they cannot depend on the result of any intermediate step of the time-evolution. 

\subsection{Turing-completeness proof}
\begin{theorem}
\label{theorem1}
For every Turing Machine $\mathcal{T}$, there exists a Degenerate Optical Parametric Oscillator Network $\mathcal{N}_{\mathcal{T}}$ with $N=12$ optical pulses that simulates $\mathcal{T}$.
\end{theorem}
\noindent\textit{Proof}:\indent We will show by explicit construction how a DOPON $\mathcal{N}_{\mathcal{T}}$ with just $N=12$ optical pulses can be used to simulate any TM. We denote these pulse amplitudes in order as $\{q,r,l,\mathbb{1},a_{1},a_{2},a_{3},a_{4},a_{5},a_{6},a_{7},a_{8}\}$ and begin by describing how the TM is encoded in $\mathcal{N}_{\mathcal{T}}$. Suppose that $\mathcal{T}$ has states $Q=\{q_{1},\ldots,q_{m}\}$, where we index the states from $1$ to $m$. If $\mathcal{T}$ is in state $q_{i}\in Q$ after step $T$ of the TM operation, then we define $\mathcal{U}:Q\rightarrow\mathbb{R}$ and require
\begin{equation}
    q(8mT)=\mathcal{U}(q_{i})\coloneqq\sum_{k=1}^{i}\frac{(-1)^{k+1}}{2^{k}}
    \label{eq:3}
\end{equation}
In other words, the TM state is represented using a unary-like encoding scheme $\mathcal{U}$ for the optical pulse amplitude $q$ in $\mathcal{N}_{\mathcal{T}}$. The idea is to choose the dynamics of $\mathcal{N}_{\mathcal{T}}$ such that the states are decoded one at a time, checking one state every 8 time steps and performing the appropriate state update.

Next, we describe how the TM tape is encoded in $\mathcal{N}_{\mathcal{T}}$. Consider the tape after step $T$ as a sequence of binary symbols $\ldots s_{-2}s_{-1}\boxed{s_{0}}s_{1}s_{2}\ldots$ where the boxed symbol $\boxed{s_{0}}$ denotes the current symbol being read by the head. We represent the tape in $\mathcal{N}_{\mathcal{T}}$ using two optical pulse amplitudes $r$ and $l$, which encode the right/left sequences $\boxed{s_{0}}s_{1}s_{2}\ldots$ and $s_{-1}s_{-2}\ldots$, respectively, and define $\mathcal{C}^{\pm}:\mathbb{N}\rightarrow\mathbb{R}$ such that
\begin{subequations}
\label{eq:4}
\begin{equation}
    r(8mT)=\mathcal{C}^{+}(s_{0}s_{1}s_{2}\ldots)\coloneqq\sum_{k=0}^{\infty}(-1)^{k}\frac{2s_{k}+1}{4^{k+1}}
\label{eq:4a}
\end{equation}
\begin{equation}
    l(8mT)=\mathcal{C}^{-}(s_{-1}s_{-2}\ldots)\coloneqq\sum_{k=1}^{\infty}(-1)^{k+1}\frac{2s_{-k}+1}{4^{k}}
\label{eq:4b}
\end{equation}
\end{subequations}

\begin{table*}[t]
\begin{ruledtabular}
\begin{tabular}{cc}
Turing Machine $\mathcal{T}$ & Degenerate Optical Parametric Oscillator Network $\mathcal{N}_{\mathcal{T}}$\\
\colrule
\begin{tabular}{@{}c@{}}State \\ $q_{i}\in Q$\end{tabular} & \begin{tabular}{@{}c@{}}Optical pulse amplitude with unary-like encoding\\ $q=\mathcal{U}(q_{i})=\sum_{k=1}^{i}\frac{(-1)^{k+1}}{2^{k}}$\end{tabular}\\
\colrule
\begin{tabular}{@{}c@{}}Tape \\ $\ldots s_{-2}s_{-1}\boxed{s_{0}}s_{1}s_{2}\ldots$\end{tabular} & \begin{tabular}{@{}c@{}}Optical pulse amplitudes with Cantor-like encoding\\ $r=\mathcal{C}^{+}(s_{0}s_{1}s_{2}\ldots)=\sum_{k=0}^{\infty}(-1)^{k}\frac{2s_{k}+1}{4^{k+1}}$\\ $l=\mathcal{C}^{-}(s_{-1}s_{-2}\ldots)=\sum_{k=1}^{\infty}(-1)^{k+1}\frac{2s_{-k}+1}{4^{k}}$\end{tabular}\\
\colrule
\begin{tabular}{@{}c@{}}Time Steps \\ $T:\mathbb{N}\rightarrow\mathbb{N}$\end{tabular} & \begin{tabular}{@{}c@{}}Optical resonator roundtrips\\ $t=8mT$\end{tabular}\\
\end{tabular}
\end{ruledtabular}
\caption{Encoding of Turing Machine $\mathcal{T}$ in Degenerate Optical Parametric Oscillator Network $\mathcal{N}_{\mathcal{T}}$}
\label{tab:1} 
\end{table*}

\begin{figure}[b]
\includegraphics[width=\linewidth]{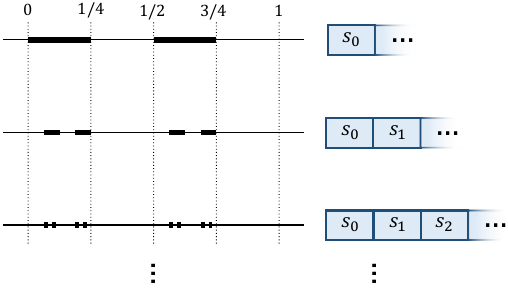}
\caption{\textbf{Cantor-like tape encoding.} The Cantor-like encoding in base 4 introduces gaps into the unit interval such that the most-significant bit of the tape can be read in constant $\mathcal{O}(1)$ time without traversing the entire tape sequence.}
\label{fig:2}
\end{figure}

This particular Cantor-like encoding scheme $\mathcal{C}^{\pm}$ in base 4 (similar to in Ref.~\cite{siegelmann1992computational}) is chosen instead of a simpler unary or binary encoding for two main reasons. Firstly, it ensures that the necessary amplitudes are bounded despite a potentially unbounded TM tape length, which avoids the unphysical scenario of an optical pulse with unbounded amplitude and energy. Secondly, consider the two numbers $0.1000\ldots$ and $0.0111\ldots$ using a binary encoding. This runs into the difficulty that distinguishing the two numbers requires traversing the entire sequence of bits, which may be potentially unbounded if representing a TM tape. To overcome this difficulty, the Cantor-like encoding $\mathcal{C}^{\pm}$ introduces `gaps' into the unit interval, as shown in Fig.~\ref{fig:2}, such that the most-significant bit can be determined in $\mathcal{O}(1)$ time. For example, using the nonlinearity available in $\mathcal{N}_{\mathcal{T}}$ as a decision threshold, we have that $\rho(8r-3)=1$ if $s_{0}=1$, and $\rho(8r-3)=-1$ if $s_{0}=0$. 

The pulse amplitude $\mathbb{1}$ is used as a constant bias or reference amplitude, and does not change from resonator roundtrip to roundtrip. Finally, the remaining pulses $a_{1},a_{2},\ldots,a_{8}$ are used as ancillary variables to help store intermediate results of the computation. The encoding of $\mathcal{T}$ in $\mathcal{N}_{\mathcal{T}}$ is summarized in Table~\ref{tab:1}.

Now, we give the full construction for the simulation of $\mathcal{T}$ by $\mathcal{N}_{\mathcal{T}}$ according to the encoding scheme defined in Table~\ref{tab:1}. The objective is to find the set of coupling weights $\{J_{ij}(t)\}$ for $i,j=1,2,\ldots,12$ and $t\in\mathbb{N}$, and the initial optical pulse amplitudes $\{q(0),r(0),l(0),\mathbb{1}(0),a_{1}(0),a_{2}(0),\ldots,a_{8}(0)\}$ that ensure the time-evolution of $\mathcal{N}_{\mathcal{T}}$ exactly reproduces the operation of $\mathcal{T}$. The initial conditions are easy, we simply let $q(0)$, $r(0)$, and $l(0)$ be as defined in Table~\ref{tab:1} for the initial configuration of $\mathcal{T}$. The bias pulse satisfies $\mathbb{1}(t)=1$ for all $t\in\mathbb{N}$, and let $a_{j}(0)=0$ for $j=1,2,\ldots,8$. The ancillary variables begin and end with zero amplitude during each step of $\mathcal{T}$. For the coupling weights $J_{ij}(t)$, the general strategy is to find a set of time-periodic weights that encode the behaviour of $\mathcal{T}$. During each step of $\mathcal{T}$, the current state is decoded in $\mathcal{N}_{\mathcal{T}}$ by iterating through all states, one state at a time. We can perform this process using 8 time steps per state, hence requiring $8m$ time steps per step of $\mathcal{T}$. Therefore, the coupling weights $J_{ij}(t)$ repeat with a period of $8m$. During each 8-cycle in $\mathcal{N}_{\mathcal{T}}$, the current symbol $s_{0}$ is also read, and the appropriate updates (if any) to $q$, $r$, and $l$ are performed. We show the construction for one 8-cycle, and the rest follows analogously. 

The first step is to check the current state. During the first 8-cycle, we check if $q=\mathcal{U}(q_{1})$, otherwise move on to the next state. This is done by reading the most-significant bit of $q$, then storing the result in $a_{3}$ through the steps:
\begin{subequations}
\label{eqs:5}
\begin{equation}
    q(1)=\rho(q)-3q+\mathbb{1}
    \label{eq:5a}
\end{equation}
\begin{equation}
    a_3(2)=\rho(a_{3})-4q
    \label{eq:5b}
\end{equation}
\begin{equation}
    a_{3}(3)=\rho(a_{3})+\mathbb{1}
\end{equation}
\end{subequations}
where we henceforth suppress the time step $t$ for optical pulse amplitudes on the right-hand side of equations since it is always one less than on the left-hand side. We use the fact that $\rho(q)=q$ since $1/4\leq q\leq 1/2$ so that Eq.~\ref{eq:5a} performs the action $q\rightarrow-2q+1$, which has the effect of subtracting the most-significant bit of $q$ and shifting all remaining bits to the left. After this action, $a_{3}$ performs a decision threshold such that $a_{3}(3)=1$ if $q=0$ and $a_{3}(3)=0$ otherwise. Therefore, we will only have $a_{3}(3)=1$ if $q(0)=\mathcal{U}(q_1)$, i.e. $q$ encoded for the state $q_{1}$ being checked during the first 8-cycle. Iterating through all states by repeating this procedure $m$ times guarantees that $a_{3}(3)=1$ during exactly one 8-cycle, which allows the current state to be effectively decoded. In parallel, we also decode the most-significant bits of $r$ and $l$, which represent the current symbol $s_{0}$ read by the head and the left-adjacent symbol $s_{-1}$, respectively:
\begin{subequations}
\label{eqs:6}
\begin{equation}
    a_{1}(1)=\rho(a_{1})+8r-3\mathbb{1}
\end{equation}
\begin{equation}
    a_{1}(2)=\rho(a_{1})
\end{equation}
\begin{equation}
    a_{1}(3)=\rho(a_{1})-\frac{1}{2}a_{1}+\frac{1}{2}\mathbb{1}
\end{equation} 
\begin{equation}
    a_{2}(1)=\rho(a_{2})+8l-3\mathbb{1}
\end{equation}
\begin{equation}
    a_{2}(2)=\rho(a_{2})
\end{equation}
\begin{equation}
    a_{2}(3)=\rho(a_{2})-\frac{1}{2}a_{2}+\frac{1}{2}\mathbb{1}
\end{equation}
\end{subequations}

Therefore, we have that $a_{1}(3)=s_{0}$ and similarly $a_{2}(3)=s_{-1}$. Next, we consider the product $a_{3}(3)a_{1}(3)$, which is needed to represent the domain $Q\times S$ over which the TM (partial) functions $G$, $F$, and $D$ are defined. We have that $a_{3}(3)a_{1}(3)=1$ if $q(0)=\mathcal{U}(q_{1})$ \textit{and} $s_{0}=1$. Similarly, $a_{3}(3)(1-a_{1}(3))=1$ if $q(0)=\mathcal{U}(q_{1})$ \textit{and} $s_{0}=0$. To construct the product pairs using only linear couplings, we make use of the identity $a\cdot b=\rho(a+b-2)+1$ for $a,b\in\{0,1\}$. We store these product pairs in $a_{5}$ and $a_{6}$, respectively. We also store a copy of $a_{3}(3)$ in $a_{7}$ for later use:
\begin{subequations}
\label{eqs:7}
\begin{equation}
    a_{5}(4)=\rho(a_{5})+a_{1}+a_{3}-2\mathbb{1}
\end{equation}
\begin{equation}
    a_{5}(5)=\rho(a_{5})+\mathbb{1}
\end{equation}
\begin{equation}
    a_{6}(4)=\rho(a_{6})-a_{1}+a_{3}-\mathbb{1}
\end{equation}
\begin{equation}
    a_{6}(5)=\rho(a_{6})+\mathbb{1}
\end{equation}
\begin{equation}
    a_{7}(5)=\rho(a_{7})+a_{3}
\end{equation}
\end{subequations}

Note that if $a_{5}(5)=1$, then $a_{6}(5)=0$, and vice versa. After reading the current state and symbol, we perform state, tape, and head updates (if any) defined by $\mathcal{T}$. First, consider the tape and head updates. Suppose that $\mathcal{T}$ is defined such that $F(q_{1},s_{0})=s^{\prime}$. If $D(q_{1},s_{0})=0$ (move head left), then the tape updates from $\ldots s_{-2}s_{-1}\boxed{s_{0}}s_{1}s_{2}\ldots$ to $\ldots s_{-2}\boxed{s_{-1}}s^{\prime}s_{1}s_{2}\ldots$ where the head is now reading symbol $\boxed{s_{-1}}$. To represent this tape change in $\mathcal{N}_{\mathcal{T}}$, the desired actions are:
\begin{subequations}
\label{eqs:8}
\begin{equation}
    l\rightarrow\mathcal{C}^{-}(s_{-2}s_{-3}\ldots)=-4l+(2s_{-1}+1)
\end{equation}
\begin{eqnarray}
    r\rightarrow\mathcal{C}^{+}(s_{-1}s^{\prime}s_{2}\ldots)&=-\frac{r}{4}+\frac{2s_{-1}+1}{4}\nonumber\\
    &+\frac{2s_{0}+1}{4^{2}}-\frac{2s^{\prime}+1}{4^{2}}
\end{eqnarray}
\end{subequations}

On the other hand, if $D(q_{1},s_{0})=1$ (move head right), then the tape updates from $\ldots s_{-2}s_{-1}\boxed{s_{0}}s_{1}s_{2}\ldots$ to $\ldots s_{-2}s_{-1}s^{\prime}\boxed{s_{1}}s_{2}\ldots$ where the head is now reading symbol $\boxed{s_{1}}$. To represent this tape change in $\mathcal{N}_{\mathcal{T}}$, the desired actions are:
\begin{subequations}
\label{eqs:9}
\begin{equation}
    l\rightarrow\mathcal{C}^{-}(s^{\prime}s_{-1}s_{-2}\ldots)=-\frac{l}{4}+\frac{2s^{\prime}+1}{4}
\end{equation}
\begin{equation}
    r\rightarrow\mathcal{C}^{+}(s_{1}s_{2}\ldots)=-4r+(2s_{0}+1)
\end{equation}
\end{subequations}

Consider the required update for $l$. We use the fact that $\rho(l)=l$ since $l\in[0,3/4]$, so the eventual desired update can be expressed as:
\begin{eqnarray}
    &l(8)=\rho(l)+4a_{5}(5)\left[\left(1-D(q_{1},1)\right)\frac{-5l+2a_{2}(3)+1}{4}\right.\nonumber\\
    &\left.+\frac{D(q_{1},1)}{4}\frac{-5l+2F(q_{1},1)+1}{4})\right]+4a_{6}(5)\Big[\left(1-D(q_{1},0)\right)\nonumber\\
    &\left.\frac{-5l+2a_{2}(3)+1}{4}+\frac{D(q_{1},0)}{4}\frac{-5l+2F(q_{1},0)+1}{4}\right]
\label{eq:10}
\end{eqnarray}

Recall that if $a_{5}(5)=1$, then $a_{6}(5)=0$, and vice versa. Thus only one of the coupling terms involving box brackets in Eq.~\ref{eq:10} can be non-zero. To re-write Eq.~\ref{eq:10} using only linear coupling terms allowed in $\mathcal{N}_{\mathcal{T}}$, we use the identity $a\cdot x=\rho(x+2a-2)+1-a$ for $a\in\{0,1\}$ and $x\in[-3/4,3/4]$. We compute the two terms involving box brackets in Eq.~\ref{eq:10} separately, and store them in $a_{3}$ (recycling an ancillary variable) and $a_{4}$:
\begin{subequations}
\label{eqs:11}
\begin{eqnarray}
    a_{3}(6)&=\rho(a_{3})-a_{3}+\left[\frac{-5(1-D(q_{1},1))}{4}-\frac{5D(q_{1},1)}{16}\right]l\nonumber\\
    &+\frac{1-D(q_{1},1)}{2}a_{2}+2a_{5}+\left[\frac{1-D(q_{1},1)}{4}\right.\\
    &\left.+\frac{D(q_{1},1)F(q_{1},1)}{8}+\frac{D(q_{1},1)}{16}-2\right]\mathbb{1}\nonumber
\end{eqnarray}
\begin{equation}
    a_{3}(7)=\rho(a_{3})-a_{5}+\mathbb{1}
\end{equation}
\begin{eqnarray}
    a_{4}(6)&=\rho(a_{4})+\left[\frac{-5(1-D(q_{1},0))}{4}-\frac{5D(q_{1},0)}{16}\right]l\nonumber\\
    &+\frac{1-D(q_{1},0)}{2}a_{2}+2a_{6}+\left[\frac{1-D(q_{1},0)}{4}\right.\\
    &\left.+\frac{D(q_{1},0)F(q_{1},0)}{8}+\frac{D(q_{1},0)}{16}-2\right]\mathbb{1}\nonumber
\end{eqnarray}
\begin{equation}
    a_{4}(7)=\rho(a_{4})-a_{6}+\mathbb{1}
\end{equation}
\end{subequations}

Therefore, the desired update for $l$ using only linear couplings is:
\begin{eqnarray}
    l(8)=\rho(l)+4a_{3}+4a_{4}\ .
\end{eqnarray}

Now, consider the required update for $r$. We use the fact that $\rho(r)=r$ since $r\in[0,3/4]$, so the eventual desired update can be expressed as:
\begin{eqnarray}
    &r(8)=\rho(r)+4a_{5}(5)\left[\frac{1-D(q_{1},1)}{4}\left(\frac{-5r}{4}+\frac{2a_{1}(3)+1}{16}\right.\right.\nonumber\\
    &\left.\left.-\frac{2F(q_{1},1)+1}{16}+\frac{2a_{2}(3)+1}{4}\right)+D(q_{1},1)\frac{-5r+2a_{1}(3)+1}{4}\right]\nonumber\\
    &+4a_{6}(5)\left[\frac{1-D(q_{1},0)}{4}\left(\frac{-5r}{4}+\frac{2a_{1}(3)+1}{16}-\frac{2F(q_{1},0)+1}{16}\right.\right.\nonumber\\
    &\left.\left.+\frac{2a_{2}(3)+1}{4}\right)+D(q_{1},0)\frac{-5r+2a_{1}(3)+1}{4}\right]
    \label{eq:13}
\end{eqnarray}
Similar to above for $l$, we compute the two terms involving box brackets in Eq.~\ref{eq:13} separately, and store them in $a_{1}$ and $a_{2}$ (recycling ancillary variables):
\begin{subequations}
\label{eqs:14}
\begin{eqnarray}
    &a_{1}(6)=\rho(a_{1})+\left(\frac{-5(1-D(q_{1},1))}{16}-\frac{5D(q_{1},1)}{4}\right)r\\
    &+\left(\frac{1-D(q_{1},1)}{32}+\frac{D(q_{1},1)}{2}-1\right)a_{1}+\frac{1-D(q_{1},1)}{8}a_{2}+2a_{5}\nonumber\\
    &+\left(\frac{1-D(q_{1},1)}{16}-\frac{(1-D(q_{1},1))F(q_{1},1)}{32}+\frac{D(q_{1},1)}{4}-2\right)\mathbb{1}\nonumber
\end{eqnarray}
\begin{equation}
    a_{1}(7)=\rho(a_{1})-a_{5}+\mathbb{1}\ ,
\end{equation}
\begin{eqnarray}
    &a_{2}(6)=\rho(a_{2})+\left(\frac{-5(1-D(q_{1},0))}{16}-\frac{5D(q_{1},0)}{4}\right)r\\
    &+\left(\frac{1-D(q_{1},0)}{32}+\frac{D(q_{1},0)}{2}\right)a_{1}+\left(\frac{1-D(q_{1},0)}{8}-1\right)a_{2}+2a_{6}\nonumber\\
    &+\left(\frac{1-D(q_{1},0)}{16}-\frac{(1-D(q_{1},0))F(q_{1},0)}{32}+\frac{D(q_{1},0)}{4}-2\right)\mathbb{1}\nonumber
\end{eqnarray}
\begin{equation}
    a_{2}(7)=\rho(a_{2})-a_{6}+\mathbb{1}\ .
\end{equation}
\end{subequations}

Therefore, the desired update for $r$ using only linear couplings is:
\begin{eqnarray}
    r(8)=\rho(r)+4a_{1}+4a_{2}\ .
\end{eqnarray}

Next, consider the state update for $q$. During the first 8-cycle, if $q(0)=\mathcal{U}(q_{1})$, then the desired action $q\rightarrow\mathcal{U}(G(q_{1},s_{0}))$. Otherwise, we simply move on to decoding for the next state $q_{2}$. This can be implemented in $\mathcal{N}_{\mathcal{T}}$ as:
\begin{eqnarray}
    q(8)&=\rho(q)-a_{3}(3)q\qquad\qquad\qquad\qquad\qquad\qquad\qquad\\
    &+\left[\sum_{k=1}^{m-1}\frac{(-1)^{k+1}}{2^{k}}+\frac{(-1)^{m}}{2^{m-1}}\mathcal{U}(G(q_{1},1))\right]a_{5}\nonumber\\
    &+\left[\sum_{k=1}^{m-1}\frac{(-1)^{k+1}}{2^{k}}+\frac{(-1)^{m}}{2^{m-1}}\mathcal{U}(G(q_{1},0))\right]a_{6}\nonumber
\end{eqnarray}

The optical pulse amplitude $q$ will correctly encode the next state $G(q_{1},s_{0})$ after repeating the analogous 8-cycle another $m-1$ times. An additional $m-1$ dummy ones are added to the front of $q$ since the most-significant bit is deleted during each 8-cycle. We use the same identity as above to express the product $a_{3}(3)q$ in terms of linear couplings stored in ancillary variable $a_{8}$:
\begin{subequations}
\begin{equation}
    a_{8}(6)=\rho(a_{8})+q+2a_{7}-2\mathbb{1}
\end{equation}
\begin{equation}
    a_{8}(7)=\rho(a_{8})-a_{7}+\mathbb{1}
\end{equation}
\end{subequations}

Therefore, the desired update for $q$ using only linear couplings is:
\begin{eqnarray}
    q(8)&=\rho(q)-a_{8}\qquad\qquad\qquad\qquad\qquad\qquad\qquad\\
    &+\left[\sum_{k=1}^{m-1}\frac{(-1)^{k+1}}{2^{k}}+\frac{(-1)^{m}}{2^{m-1}}\mathcal{U}(G(q_{1},1))\right]a_{5}\nonumber\\
    &+\left[\sum_{k=1}^{m-1}\frac{(-1)^{k+1}}{2^{k}}+\frac{(-1)^{m}}{2^{m-1}}\mathcal{U}(G(q_{1},0))\right]a_{6}\nonumber
\end{eqnarray}

More generally, during the $i^{\mathrm{th}}$ 8-cycle corresponding to the first step of $\mathcal{T}$, if $q(0)=\mathcal{U}(q_{i})$, then we have:
\begin{eqnarray}
    q(8i)&=\rho(q)-a_{8}\qquad\qquad\qquad\qquad\qquad\qquad\qquad\\
    &+\left[\sum_{k=1}^{m-i}\frac{(-1)^{k+1}}{2^{k}}+\frac{(-1)^{m-i+1}}{2^{m-i}}\mathcal{U}(G(q_{i},1))\right]a_{5}\nonumber\\
    &+\left[\sum_{k=1}^{m-i}\frac{(-1)^{k+1}}{2^{k}}+\frac{(-1)^{m-i+1}}{2^{m-i}}\mathcal{U}(G(q_{i},0))\right]a_{6}\nonumber
\end{eqnarray}

Finally, the last step is to reset the ancillary variables during each 8-cycle:
\begin{equation}
    a_{j}(8)=\rho(a_{j})-a_{j}\quad \mathrm{for}\ j=1,2,\ldots,8\ .
    \label{eq:18}
\end{equation}

\begin{figure*}[t]
\includegraphics[width=\linewidth]{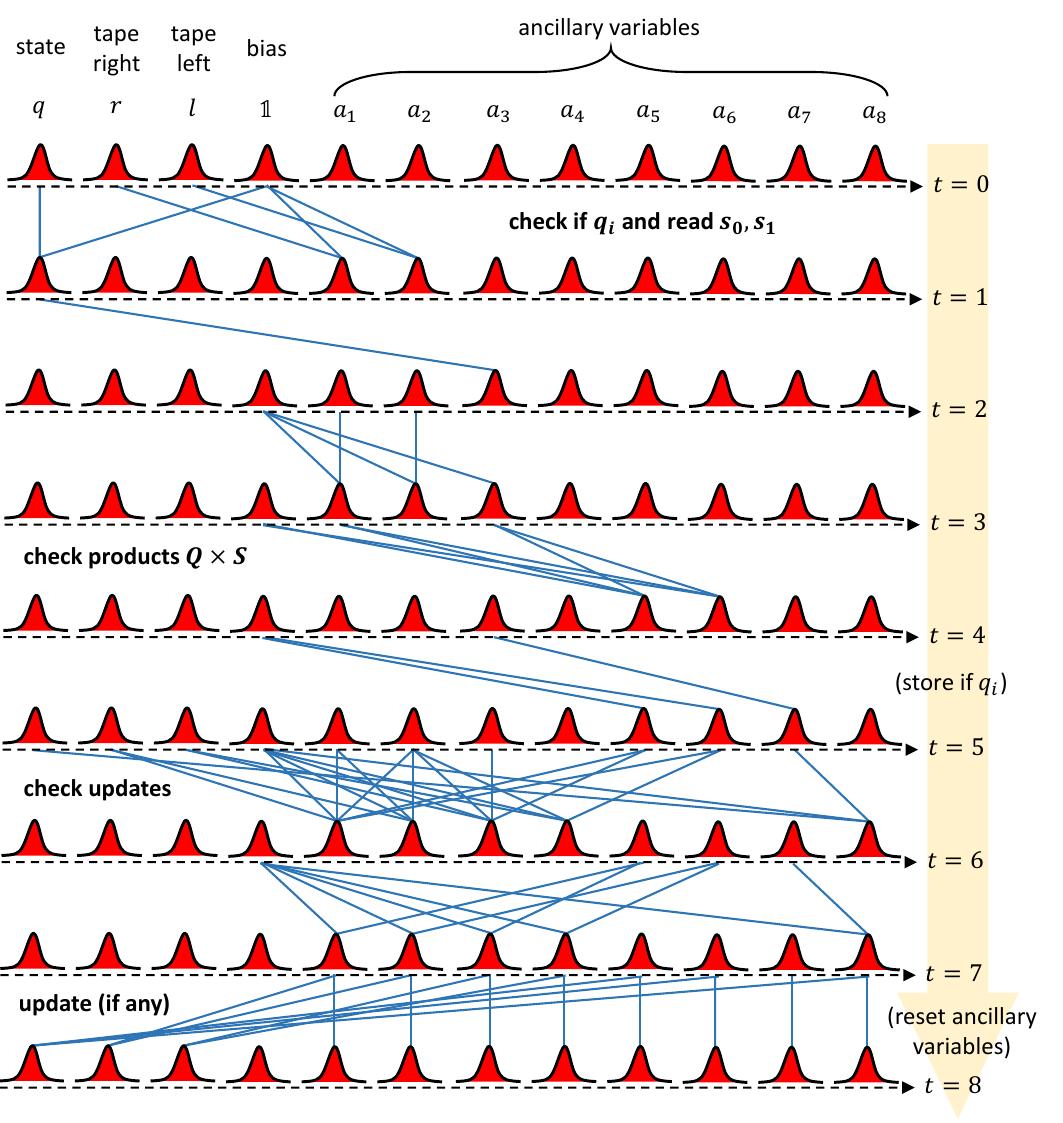}
\caption{\textbf{Turing Machine simulation steps.} Network connectivity for a DOPON simulating a TM where red pulses represent the time-multiplexed laser pulses in the DOPON during each time step $t\equiv0,1,2,\ldots,7\ (\mathrm{mod}\ 8)$ and blue lines show the active pulse couplings $J_{ij}(t)\neq 0$ during each 8-cycle. The 8-cycle repeats $m$ times per step of the TM where $m$ is the number of TM finite control states, so the coupling weights $J_{ij}(t)$ repeat with a period of $8m$ to exactly simulate the TM.}
\label{fig:3}
\end{figure*}

The construction is complete. The coupling weights for the proceeding 8-cycles follows analogously to the first 8-cycle with just minor modifications to the coupling weights according to the definition of $\mathcal{T}$ via the (partial) functions $G$, $F$, and $D$. A visual summary of the construction showing the pulses actively involved during the time steps of each 8-cycle is shown in Fig.~\ref{fig:3}.$\quad\quad\quad\quad\quad\square$

\subsection{Physical consequences}
The Turing-completeness of DOPONs has many physical implications that result from the undecidability of the \textit{Entscheidungsproblem} (the problem of determining whether an arbitrary program and input will either finish running or continue to run forever)~\cite{turing1937on}. Together with Theorem~\ref{theorem1}, this implies that \textit{the problem of determining if a DOPON (or CNOR more generally), with arbitrary couplings and initial conditions, will eventually reach a steady-state or periodic oscillation is undecidable}. This is because we can construct a DOPON that simulates a Universal Turing Machine according to the scheme used to prove Theorem.~\ref{theorem1}. Therefore, halting in the TM corresponds to a steady-state/periodic oscillation in the corresponding DOPON. 

In fact, any decision problem that can ultimately be reduced to the existence of a steady-state/periodic oscillation in DOPONs is also undecidable. For example, DOPONs have been used extensively as physical Ising solvers for combinatorial optimization~\cite{wang2013coherent}. Previous numerical studies~\cite{reifenstein2021coherent} have used a time-to-solution to quantify the problem run-time. Notably, this usually involves \textit{ad hoc} disregarding any instance in which the DOPON exceeds some finite cut-off time to reach a steady-state representing the solution to the combinatorial optimization problem. Imposing such a finite cut-off is of course necessary for practical reasons. However, the existence of long-running DOPONs is not just a numerical artifact or bug, but rather it is a feature of their Turing-completeness and moreover \textit{the time-to-solution in DOPONs is uncomputable}. Suppose that there exists a finite time-to-solution $T$ for any DOPON $\mathcal{N}$. Then we can say that $\mathcal{N}$ will be in a steady-state/periodic oscillation after a time of $T$. But, this implies that the Halting Problem for $\mathcal{N}$ is decidable. This is a contradiction, as discussed above. Therefore, there does not exist any procedure to always find such a finite $T$. 

A related problem is determining if a DOPON will find a local minimum solution or if it will find the desired global minimum solution to an optimization problem. There have been efforts to develop heuristics and scaling laws based on analyses of DOPONs with small $N$ that are tractable, then extrapolating the results to better understand problems with large $N$~\cite{ng2022efficient}. However, this strategy is not prudent because our results imply that there exists a finite $N$ below which the DOPON dynamics are decidable, but above which the DOPON dynamics are undecidable. Therefore, \textit{there is no way to decide if heuristics developed for small $N$ will also be valid for large $N$}. In fact, in our DOPON model, this threshold is as low as $N=12$. A caveat is that the heuristics are often not seeking to guarantee optimal solutions, and it may still be possible to develop heuristics that yield near-optimal solutions within some error range of the optimal solution. 

\section{Discussion}
Since undecidability is not often discussed in the context of optics, we emphasize that undecidability is a property regarding infinite classes of functions. Specific instances of a problem class are of course solvable. In addition, the simpler the system under consideration, the more powerful the result since adding complexity generally only increases the computational power of a system. Therefore, we consider a simple class of CNORs based on DOPONs so that our results are as widely-applicable as possible. We showed that a DOPON with as few as $N=12$ pulses is already sufficient for Turing-completeness. This is well-within current experimental capabilities, which means our results have practical consequences. However, we do not claim that our construction is optimal. This naturally begs the question: what is the minimum $N$ needed for Turing-completeness?

There are several obvious methods to further reduce the necessary $N$. Our construction uses a DOPON that efficiently simulates a TM since the number of time steps is linear $\mathcal{O}(T)$. But, using a less efficient TM simulation with an exponential slowdown, e.g. based on Minsky Register Machines~\cite{minsky1967computation}, can allow one to decrease $N$ at the expense of a worse time-complexity. Another way to reduce $N$ is to use more complicated dynamics. For example, the coupled cavities implementation of DOPONs shown in Fig.~\ref{fig:1}c actually allows more general dynamics than stated in Eq.~\ref{eq:1}. It can be shown that a coupled cavity design with a main cavity containing $N=6$ pulses and memory cavity containing $N+1=7$ pulses is Turing-complete. The present construction can also be modified for DOPONs with nonlinear function $\rho$ different to the simple saturable gain~\cite{li2023all}. It may also be interesting to investigate the effects of different constraints on the computational power of CNORs. For example, many experimental CNORs utilize conservative, static, and nearest-neighbour couplings. In contrast, our construction uses dissipative, time-varying, and all-to-all couplings. We expect that the minimum $N$ needed for Turing-completeness will increase greatly if conservative, static, or nearest-neighbour coupling constraints are imposed. 

In any Turing-complete model of computation, there is always the issue of infinity due to the potentially unbounded length of the TM tape, which may be considered unphysical. In our DOPON model, the number of physical components, e.g. waveguides and modulators, is constant. Moreover, the number of pulses $N$ is constant and the pulse amplitudes are bounded. The hidden infinity in our DOPON model is therefore the potentially unbounded precision needed for specifying the pulse amplitudes $l$ and $r$ that represent the TM tape. Interestingly, there is nothing obvious in classical optics that fundamentally forbids this potentially unbounded precision. Furthermore, in our construction, the coupling weights $J_{ij}(t)$ are purely rational numbers and have a \textit{finite} maximum precision. The need for potentially unbounded precision in Turing-complete DOPONs should not be confused with chaos, which results from a sensitivity to initial conditions. Our results show that even with perfect knowledge of the initial conditions, the dynamics of DOPONs is still unpredictable. 

On the other hand, in quantum optics, it is well-known that DOPONs have quantum-noise dynamics, which is often cited as a computational resource for escaping local minima~\cite{yamamoto2017coherent}. In this case, there are quantum limits to the maximum precision and signal-to-noise ratio when measuring optical pulse amplitudes. Our simple model for DOPONs operating deterministically above threshold breaks down when considering DOPONs operating stochastically with quantum noise below threshold, or in the case of transitioning stochastically from below to above threshold. It may be interesting to investigate the computational power of stochastic DOPONs in the context of non-deterministic or probabilistic TMs. We leave it as an open problem to study the computational power of DOPONs with finite precision in the context of finite automata, with the ultimate goal of establishing a hierarchy for the computational power of DOPONs given a certain size $N$ and/or finite amount of precision. 

In summary, we have shown that coupled nonlinear optical resonators can be used as universal computers. Our proof is based on an explicit construction using a degenerate optical parametric oscillator network, which shows that as few as only $N=12$ pulses is needed for Turing-completeness. This has profound consequences for both experiments and applications because any physical property reducible to the existence of a steady-state or periodic oscillation is logically undecidable. We hope that our findings will stimulate further inquiry into the computational properties of coupled nonlinear optical resonators and inform their use as part of next-generation optical computing systems.  

\subsection*{Acknowledgments}
The authors acknowledge support from ARO grant no. W911NF-23-1-0048, NSF grant no. 1918549, Center for Sensing to Intelligence at Caltech, and NASA/JPL. G.H.Y.L acknowledges support from the Quad Fellowship. 
\subsection*{Author Contributions}
All authors contributed to this manuscript.
\subsection*{Data Availability}
The data used to generate the plots and results in this paper are available from the corresponding author upon reasonable request.
\subsection*{Competing Interests}
AM has financial interest in PINC Technologies Inc., which is developing photonic integrated nonlinear circuits. The remaining authors declare no competing interests.
\bibliography{apssamp}
\end{document}